\documentclass[10pt,twocolumn,journal]{IEEEtran}

\usepackage{psfrag}
\usepackage{graphicx}
\usepackage[english]{babel}
\usepackage{amssymb}
\usepackage{amsbsy}
\usepackage{multirow}
\usepackage{bigstrut}
\usepackage{slashbox}
\usepackage{colortbl}
\usepackage{threeparttable}
\usepackage{algorithm}
\usepackage{algorithmic}
\usepackage{cite}

\title{Improving Network-on-Chip-based turbo decoder architectures}

\author{Maurizio Martina, \emph{Member IEEE}, Guido Masera, \emph{Senior Member IEEE} 
\thanks{The authors are with 
Dipartimento di Elettronica - Politecnico di Torino - Italy. 
}}

\begin{document}

\maketitle

\begin{abstract}
In this work novel results concerning Network-on-Chip-based turbo decoder architectures are presented.
Stemming from previous publications, this work concentrates first on improving the throughput by exploiting 
adaptive-bandwidth-reduction techniques. 
This technique shows in the best case an improvement of more than 60 Mb/s. 
Moreover, it is known that double-binary turbo decoders require higher 
area than binary ones. This characteristic has the negative effect of increasing the data width of the network nodes. 
Thus, the second contribution of this work is to reduce the network complexity to support double-binary codes, 
by exploiting bit-level and pseudo-floating-point representation of the extrinsic information. 
These two techniques allow for an area reduction of up to more than the 40\% with a performance degradation of 
about 0.2 dB.
\end{abstract}


\section{Introduction}
\label{sec:intro}

Today, modern telecommunications are a pervasive experience of data exchange among users and devices. 
One critical aspect of this scenario is the continuous demand for higher data rates, a problem that is exacerbated 
by the need for reliable transmission of data.
To that purpose, the push on the so-called beyond-3G technologies, such as WiMAX \cite{802-16} and 3GPP-LTE \cite{lte}, 
is a possible answer, where the reliability is obtained exploiting effective error correcting codes, 
such as turbo \cite{berrou_ICC93} and LDPC \cite{gallager_TrIT62} codes. 
Unfortunately, the decoding algorithms for these codes are iterative making high throughput implementations 
a challenging task \cite{montorsi_IEEEProc07, boutillon_TCOM07}. 

As shown in Table I in \cite{martina_MPMS11}, several modern standards for communications use turbo codes 
as a reliable channel coding scheme. However, since these codes have limited similarities, flexible architectures 
able to support different standards are interesting solutions to achieve interoperability \cite{polydoros_PIMRC08}. 
This direction has been investigated in several works \cite{wehn_TVLSI08, bougard_ICT08, martina_TCASII08, kim_CICC09, baghdadi_TVLSI09, baghdadi_soc10, wehn_ICCS10} 
where not only flexibility but also high throughput, achieved by the means of parallel architectures, is addressed.
As an example \cite{martina_TCASII08, kim_CICC09, wehn_ICCS10} deal with optimized ASIC architectures where the 
flexibility is limited to two standards, UMTS/WiMAX, 3GPP-LTE/WiMAX and 3GPP-LTE/HSDPA respectively. 
On the other hand, \cite{wehn_TVLSI08, bougard_ICT08, baghdadi_TVLSI09, baghdadi_soc10} are based on the ASIP approach, where 
optimized processor-like architectures are used. 
It is worth observing that ASIP-based solutions allow for greater flexibility than ASIC-based architectures, as they can support several different codes and 
standards. 
Moreover, as suggested in \cite{baghdadi_TVLSI09}, ASIP solutions are well suited to implement high throughput 
multiprocessor turbo decoder architectures \cite{martina_MPMS11}. 

Recently, in \cite{vacca_DSD09} we introduced the concept of intra-IP Network-on-Chip (NoC), where the well 
known NoC paradigm is applied to the communication structure of processing elements that belong to the same IP. 
As discussed in several works, such as 
\cite{wehn_iscas05, moussa_date07, moussa_iscas08, martina_TCASI10, martina_MPMS11}, intra-IP NoC is 
a flexible solution to enable multi-ASIP turbo decoder architectures. 
However, as shown in detail in \cite{martina_TCASI10, martina_MPMS11}, flexibility comes at the expense 
of increasing the complexity of the decoder architecture. In this work we improve 
the complexity/performance trade-off of NoC-based turbo decoder architectures by reducing the traffic load 
on the network as suggested in \cite{baghdadi_EL06}. 
The adopted technique of traffic reduction offers in the best case a throughput improvement 
of more than 60 Mb/s and 40 Mb/s for binary and double-binary codes respectively. 
Furthermore, we exploit two known techniques 
\cite{lee_VTC08,kim_TCASII09}, originally proposed to limit the amount of memory in turbo decoder architectures, 
as possible solutions to reduce the complexity of the NoC when double-binary turbo codes \cite{berrou_ITW01} are employed, 
as in the WiMAX standard. 
\begin{figure}[th!]
  \centering
  \includegraphics[width=\columnwidth]{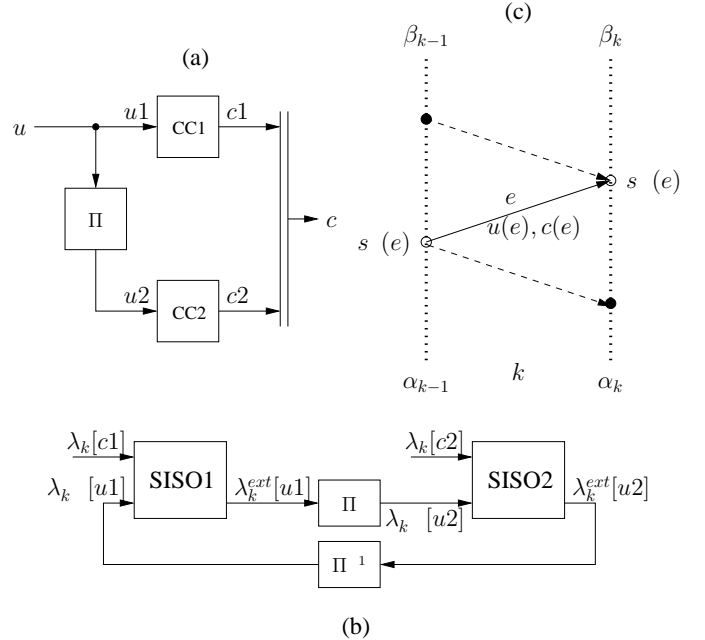}
\caption{Parallel concatenation of two convolutional codes: encoder (a), decoder (b), notation for a trellis section (c)} 
\label{fig:turbo}
\end{figure}

The paper is structured as follows: in section \ref{sec:theory} we recall the equations required to implement the 
decoding algorithm, whereas in section \ref{sec:noc} we describe the peculiar characteristics of an NoC-based turbo 
decoder architecture, including the architecture of routing elements, low-complexity routing algorithms and topologies.
Section \ref{sec:setup} describes the experimental setup we defined to increase the throughput and reduce the area 
of NoC-based turbo decoder architectures both in the case of binary and double-binary codes. To this purpose 
we considered the HSDPA and the 3GPP-LTE standards for the case of binary codes, and the WiMAX standard for the 
case of double-binary codes. Finally, in section \ref{sec:concl} conclusions are drawn.
\begin{figure*}[t!]
  \centering
  \includegraphics[width=\textwidth]{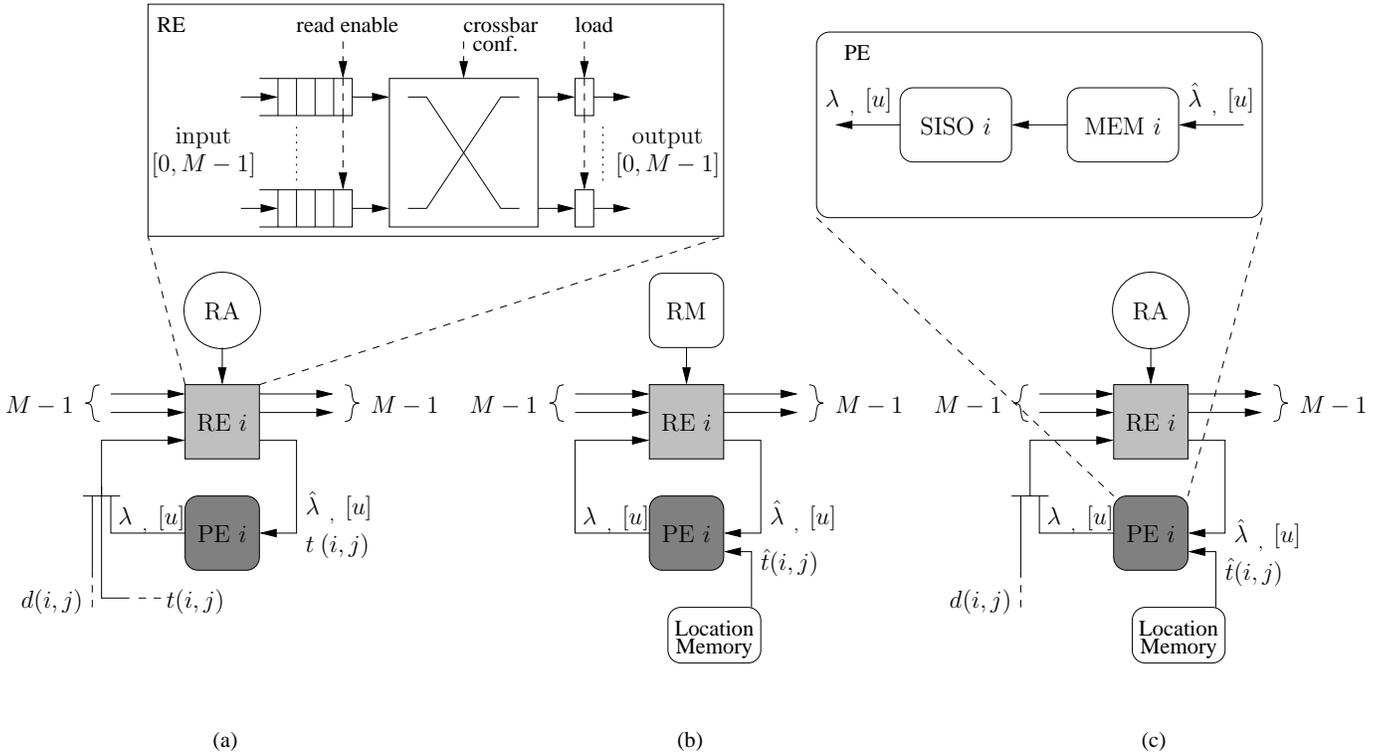}
\caption{Node block scheme: 
(a) FA architecture, 
(b) AP architecture, 
(c) PP architecture}
\label{fig:nodes}
\end{figure*}

\section{Decoding algorithms}
\label{sec:theory}

Since turbo codes are based on the concatenation (usually parallel) 
of two constituent Convolutional Codes (CC) (Fig. \ref{fig:turbo} (a)), the decoder is made of 
two constituent decoders that exchange their data by means of an interleaver ($\Pi$) and a deinterleaver ($\Pi^{-1}$), 
see Fig. \ref{fig:turbo} (b). For the sake of brevity in the next paragraph 
we define the symbols used in Fig. \ref{fig:turbo} (a) and (b) 
without specifying if they are related to CC1 or CC2.

The decoding algorithm of turbo codes is an iterative process made of two half iterations, one for each constituent 
decoder, where each half iteration is based on Maximum-A-Posteriori (MAP) estimation 
achieved by means of the BCJR algorithm \cite{bahl_TrIT94}, where Log-Likelihood-Ratio (LLR) representation is 
usually adopted \cite{robertson_ICC95}. 
Based on the trellis notation shown in Fig. \ref{fig:turbo} (c) and said $\mathcal{U}$ the set of uncoded symbols, 
each constituent MAP decoder, often referred to as Soft-In-Soft-Out (SISO) module, computes 
\begin{equation}
 \lambda^{ext}_k[u] = \max^*_{e:u(e)=u} \{ b^{ext}(e) \} - \max^*_{e:u(e)=\tilde{u}} \{ b^{ext}(e) \} - \lambda^{apr}_k[u]
\label{eq:lambda_out}
\end{equation}
where $\tilde{u} \in \mathcal{U}$ is a uncoded symbol taken as a reference (usually $\tilde{u}=\mathbf{0}$), 
$u \in \mathcal{U} \setminus \{\tilde{u}\}$, 
$k$ is a trellis step, $e$ is a transition in a trellis step and $u(e)$ is the corresponding uncoded symbol. 
Thus, $\lambda^{ext}_k[u]$ and $\lambda^{apr}_k[u]$ are extrinsic and a-priori information respectively 
for symbol $u$ at trellis step $k$ expressed as LLRs.
The $\displaystyle{\max^* \{x_i\}}$ function is implemented as 
$\max\{x_i\}$ followed by a correction term often stored in a small Look-Up-Table (LUT) 
\cite{robertson_ETT97, martina_CL09}. 
The correction term, usually adopted when decoding binary
codes (Log-MAP), can be omitted for double-binary turbo codes with minor error rate performance degradation (Max-Log-MAP). 

The term $b^{ext}(e)$ in (\ref{eq:lambda_out}) is defined as:
\begin{equation}
b^{ext}(e) = \alpha_{k-1} [s^S(e)] + \gamma^{ext}_k[e] + \beta_k[s^E(e)]  
\label{eq:be}
\end{equation}
\begin{equation}
\alpha_k[s] = \max_{e:s^E(e)=s} 
              \left\{ \alpha_{k-1}[s^S(e)] + \gamma_k[e] \right\}
\label{eq:alpha}
\end{equation}
\begin{equation}
\beta_k[s] = \max_{e:s^S(e)=s} \left\{ \beta_{k+1}[s^E(e)] + \gamma_k[e] \right\}
\label{eq:beta}
\end{equation}
\begin{eqnarray}
\gamma_k[e] & = & \lambda_k[\mathbf{u}(e)] + \lambda_k[\mathbf{c^u}(e)] + \lambda_k[\mathbf{c^p}(e)] \\
\gamma^{ext}_k[e] & = & \lambda_k[\mathbf{c^p}(e)]
\label{eq:gamma}
\end{eqnarray}
where $s^S(e)$ and $s^E(e)$ are the starting and the ending states of 
$e$, $\alpha_k[s^S(e)]$ and $\beta_k[s^E(e)]$ are 
the forward and backward metrics associated to 
$s^S(e)$ and $s^E(e)$ respectively. 
The terms $\lambda_k[\mathbf{u}(e)]$, $\lambda_k[\mathbf{c^u}(e)]$ and $\gamma^{ext}_k[\mathbf{c^p}(e)]$ are obtained 
adding the corresponding a-priori, intrinsic systematic and intrinsic parity LLRs respectively. 

In a parallel decoder $P$ SISOs operate concurrently on disjoint portions of 
the trellis. Said $N$ the number of trellis steps processed by each constituent decoder, we have that each SISO  
operates on a trellis slice made of $N/P$ steps. As a consequence, we can extend the notation introduced in the previous 
paragraph to a parallel decoder, where $\lambda^{ext}_{i,j}[u]$ is the extrinsic information produced by SISO $i$ at the  
$j$-th trellis step. 
For further details on the decoding algorithm the reader can refer to \cite{montorsi_IEEEProc07}.

\section{NoC-based turbo decoder architectures}
\label{sec:noc}

An NoC-based turbo decoder architecture can be represented as a graph where each node is 
made of a Routing Element (RE) and a Processing Elements (PE) (see Fig. \ref{fig:nodes}).
Each PE, devoted to perform the processing required 
by the BCJR algorithm, contains a SISO processor and two memories where 
intrinsic and a-priori information are stored respectively. On the other hand, each RE has a simple structure made of 
$M$ input buffers (FIFOs), an $M\times M$ crossbar switch and $M$ output registers. 
REs are devoted to route the data produced by PEs to the correct destination node according to $\Pi$ and 
$\Pi^{-1}$. To this purpose we introduce $d(i,j)$ as the destination node of $\lambda^{ext}_{i,j}[u]$. 
In order to complete a half iteration, $\lambda^{ext}_{i,j}[u]$ is stored at the location $t(i,j)$ in the 
a-priori information memory of node $d(i,j)$. 

In general PEs and REs can operate at different rates, thus, 
to decouple the design of PEs and REs we define $R$ as the number of packets injected in the network in a clock cycle. 
As a consequence, $R=1$ means that each PE injects in the network one new packet per clock cycle, whereas 
$R=0.5$ means that a new packet is injected in the network every two clock cycles. It is worth noting that 
the case $R=1$ corresponds to REs and PEs working at the same clock frequency (isochronous), 
with PEs able to output new packet of extrinsic information at each clock cycle. 
On the contrary, $R<1$ models either an isochronous system where PEs output less that one packet per clock cycle 
or a mesochronous system where REs work at a higher clock frequency that PEs. 
\begin{figure}[t!]
  \centering
  \psfrag{THR=1}[Bl][Bl][.6][0]{no ABR}
  \psfrag{THR=2}[Bl][Bl][.6][0]{$K=4$}
  \psfrag{THR=3}[Bl][Bl][.6][0]{$K=6$}
  \psfrag{THR=4}[Bl][Bl][.6][0]{$K=8$}
  \psfrag{THR=5}[Bl][Bl][.6][0]{$K=10$}
  \psfrag{THR=6}[Bl][Bl][.6][0]{$K=12$}
  \psfrag{THR=7}[Bl][Bl][.6][0]{$K=14$}
  \psfrag{THR=8}[Bl][Bl][.6][0]{$K=16$}
  \psfrag{THR=9}[Bl][Bl][.6][0]{$K=18$}
  \psfrag{THR=10}[Bl][Bl][.6][0]{$K=20$}
  \psfrag{THR=11}[Bl][Bl][.6][0]{$K=22$}
  \psfrag{THR=12}[Bl][Bl][.6][0]{$K=24$}
  \psfrag{THR=13}[Bl][Bl][.6][0]{$K=26$}
  \psfrag{THR=14}[Bl][Bl][.6][0]{$K=28$}
  \includegraphics[width=\columnwidth]{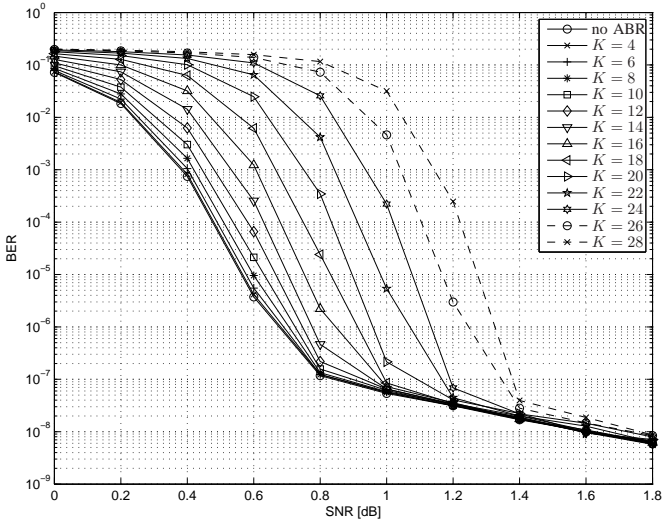}
\caption{BER performance of the HSDPA $N=5114$ turbo decoder with ABR technique for different values of $K$} 
\label{fig:hsdpa}
\end{figure}

\subsection{RE architectures}
\label{subsec:REarch}

In \cite{martina_TCASI10} three possible architectures for REs (see 
Fig. \ref{fig:nodes}), referred to as 
Fully-Adaptive (FA), All Precalculated (AP) and Partially Precalculated (PP) architectures were presented. 

The FA architecture (Fig. \ref{fig:nodes} (a)) sends on the network packets of data made of a header, containing $d(i,j)$
and a payload containing $\lambda^{ext}_{i,j}[u]$ and $t(i,j)$. The data are routed by the means of 
a Routing Algorithm (RA).

The AP architecture (Fig. \ref{fig:nodes} (b)) is obtained observing that: given $\Pi$ and $\Pi^{-1}$ we have
\begin{eqnarray}
d(i,j) & = & \left\lfloor \frac{P}{N} \cdot \Theta \left( i\cdot \frac{N}{P}+j \right) \right\rfloor \\
t(i,j) & = & \Theta\left( i\cdot \frac{N}{P}+j \right) \bmod \frac{N}{P}
\end{eqnarray}
where $\lfloor \cdot \rfloor$ is the next lowest integer value and 
$\Theta(\cdot)$ can be either $\Pi(\cdot)$ or $\Pi^{-1}(\cdot)$ depending on the current half iteration.
As a consequence, for each node we can precalculate and store in a Routing Memory (RM) and in a Location Memory 
the routing information and $\hat{t}(i,j)$, the location where the received value $\hat{\lambda}^{ext}_{i,j}[u]$ 
will be stored, respectively. Thus, with the AP architecture we reduce the width of the data bus at the expense of 
some extra memory. 

The PP architecture (Fig. \ref{fig:nodes} (c)) only precalculates the $\hat{t}(i,j)$ sequences thus, it requires a narrower data width 
than the FA architecture, but less memory than the AP one.

To improve the throughput/area figures of NoC-based turbo decoder architecture we infer from
\cite{martina_TCASI10} two main results:
\begin{itemize}
\item The AP architecture can be conveniently used with complex routing 
algorithms to concurrently maximize the throughput and minimize the area. Unfortunately, as pointed out 
in \cite{martina_MPMS11} this comes at the expense of a significant amount of external memory 
to store the routing information; as an example to support all the interleavers specified by the HSDPA 
standard \cite{hsdpa} about 64 MB of memory are required. 
\item As long as the network is faster than the PEs ($R<1$), throughput and area figures tend to be 
independent of the routing algorithm. 
\end{itemize}
Thus, both FA and PP architectures with simple RAs, should be further investigated. 
In particular, the performance of the FA architecture can be improved by using Adaptive Bandwidth Reduction (ABR) techniques 
as the one proposed in \cite{baghdadi_EL06}, namely avoiding the exchange of unnecessary extrinsic information values.
This distinguishing feature of the FA architecture, that is not available with AP and PP architectures, is detailed 
in section \ref{subsec:abr}.
On the contrary, the PP architecture features a narrower data bus than the FA one, however, it requires some external 
memory to store the configurations of all the Location Memories. 
Moreover, in several standards, such as HSDPA, 3GPP-LTE and WiMAX, the generation of $d(i,j)$ and $t(i,j)$ sequences 
can be obtained algorithmically with simple architectures \cite{kim_CICC09, wang_TCAS207, martina_CL08}. 
As a consequence, the FA architecture can also take advantage of this feature to reduce the complexity of the whole 
decoder.
\begin{figure}[t!]
  \centering
  \psfrag{THR=0}[Bl][Bl][.6][0]{no ABR}
  \psfrag{THR=4}[Bl][Bl][.6][0]{$K=4$}
  \psfrag{THR=6}[Bl][Bl][.6][0]{$K=6$}
  \psfrag{THR=8}[Bl][Bl][.6][0]{$K=8$}
  \psfrag{THR=10}[Bl][Bl][.6][0]{$K=10$}
  \includegraphics[width=\columnwidth]{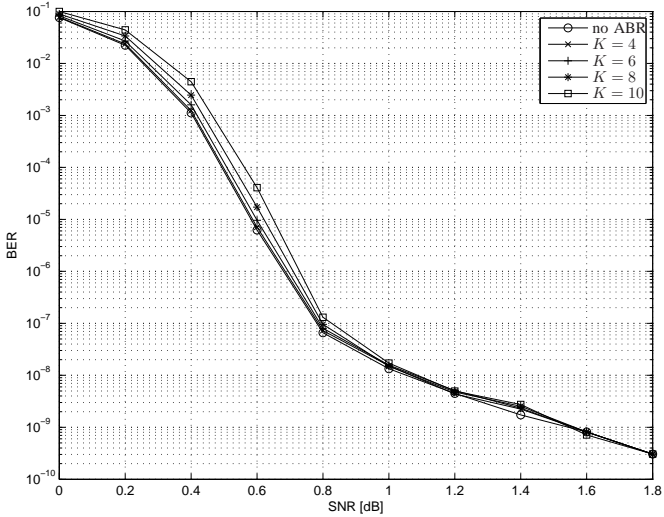}
\caption{BER performance of the 3GPP-LTE $N=6144$ turbo decoder with ABR technique, $K=4, 6, 8, 10$} 
\label{fig:lte}
\end{figure}

\subsection{Low complexity RAs}

In order to increase the throughput and reduce the area of the decoder, RAs should be based on simple, deadlock-free 
routing policies than can be implemented with few logic and completed in one clock cycle. 
As suggested in \cite{martina_TCASI10} Round-Robin (RR) and FIFO-length (FL) are suitable policies for NoC-based turbo 
decoders. 
RR is based on a circular serving policy, whereas with FL policies each input is served considering 
the number of elements stored in its input buffer, namely 
FL sorts the input buffers according to the number of stored elements, then it serves them in decreasing order.
Routing paths are stored into a routing table: for each couple of nodes in the network, one shortest-path is stored in the 
routing table. This approach, where only one shortest-path is considered, will be referred to as Single-Shortest-Path 
(SSP) \cite{martina_TCASI10} in the rest of the paper.

\subsection{NoC topologies}

In \cite{martina_TCASI10} several fixed degree topologies for NoC-based turbo decoder architectures are considered. 
However, since $\Pi$ and $\Pi^{-1}$ tend to spread almost uniformly $\lambda^{ext}_{i,j}[u]$, the 
traffic pattern on the network is almost uniform too. 
Experimental results in \cite{martina_TCASI10} show that topologies with logarithmic diameter as 
generalized De-Bruijn \cite{imase_TC81} and generalized Kautz \cite{imase_TC83} 
achieve higher throughput and require lower area than other well known fixed degree topologies such as ring, 
honeycomb and toroidal-mesh ones.
\begin{figure}[t!]
  \centering
  \psfrag{FL}[Bl][Bl][.55][0]{FL no ABR}
  \psfrag{FL THR=4}[Bl][Bl][.55][0]{FL $K=4$}
  \psfrag{FL THR=6}[Bl][Bl][.55][0]{FL $K=6$}
  \psfrag{FL THR=8}[Bl][Bl][.55][0]{FL $K=8$}
  \psfrag{FL THR=10}[Bl][Bl][.55][0]{FL $K=10$}
  \psfrag{RR}[Bl][Bl][.55][0]{RR no ABR}
  \psfrag{RR THR=4}[Bl][Bl][.55][0]{RR $K=4$}
  \psfrag{RR THR=6}[Bl][Bl][.55][0]{RR $K=6$}
  \psfrag{RR THR=8}[Bl][Bl][.55][0]{RR $K=8$}
  \psfrag{RR THR=10}[Bl][Bl][.55][0]{RR $K=10$}
  \includegraphics[width=\columnwidth]{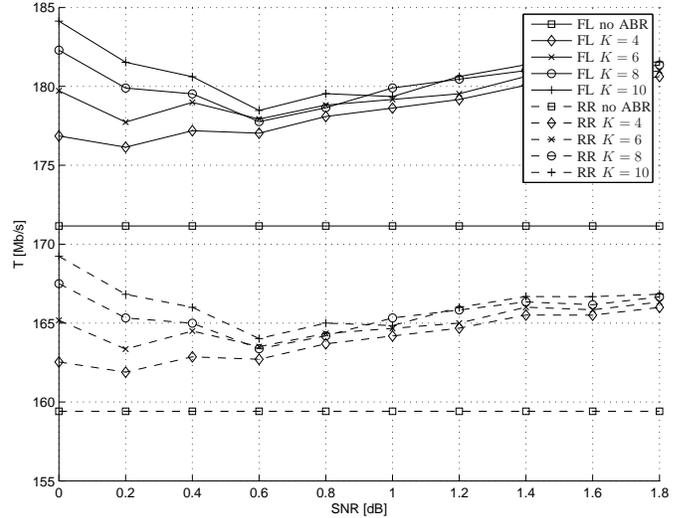}
\caption{Average throughput improvement at different SNR values of the HSDPA $N=5114$ turbo decoder with $K=4, 6, 8, 10$ 
on generalized Kautz networks $D=2$ and $P=64$}
\label{fig:ABR_hsdpa2}
\end{figure}
\begin{figure}[t!]
    \centering
      \psfrag{FL}[Bl][Bl][.55][0]{FL no ABR}
      \psfrag{FL THR=4}[Bl][Bl][.55][0]{FL $K=4$}
      \psfrag{FL THR=6}[Bl][Bl][.55][0]{FL $K=6$}
      \psfrag{FL THR=8}[Bl][Bl][.55][0]{FL $K=8$}
      \psfrag{FL THR=10}[Bl][Bl][.55][0]{FL $K=10$}
      \psfrag{RR}[Bl][Bl][.55][0]{RR no ABR}
      \psfrag{RR THR=4}[Bl][Bl][.55][0]{RR $K=4$}
      \psfrag{RR THR=6}[Bl][Bl][.55][0]{RR $K=6$}
      \psfrag{RR THR=8}[Bl][Bl][.55][0]{RR $K=8$}
      \psfrag{RR THR=10}[Bl][Bl][.55][0]{RR $K=10$}
      \includegraphics[width=\columnwidth]{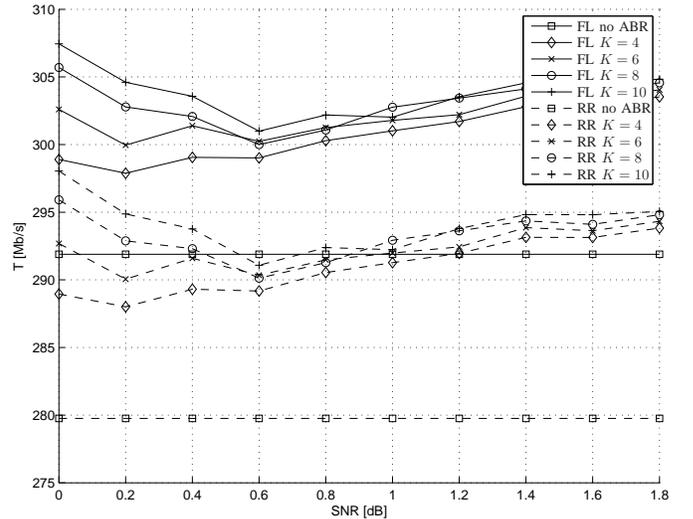}
\caption{Average throughput improvement at different SNR values of the HSDPA $N=5114$ turbo decoder with $K=4, 6, 8, 10$ 
on generalized Kautz networks $D=3$ and $P=64$}
\label{fig:ABR_hsdpa3}
\end{figure}

\section{Experimental setup}
\label{sec:setup}

Since in this work we aim at increasing the throughput and reducing the area of NoC-based turbo decoder 
architectures, we focus on the most significant cases discussed in section \ref{sec:noc}, namely FA node 
architecture with SSP-RR and SSP-FL routing algorithms. Moreover, we consider only 
generalized Kautz topologies, as they have logarithmic diameter and less self-loops\footnote{If we model a
topology as a graph, a self-loop is an edge whose source and destination nodes coincide.} than 
generalized De-Bruijn ones \cite{imase_TC81, imase_TC83, martina_TCASI10}. The degree of the network $D=M-1$ ranges in $\{2, 3, 4\}$ and the parameter $R$ varies in 
$\{ 0.33, 0.5, 1\}$. Then we simulated both HSDPA and 3GPP-LTE interleavers for the case of binary turbo 
codes. Furthermore, we simulated the double-binary turbo code used in the WiMAX standard as well.

In the following the throughput is computed as
\begin{equation}
T = \frac{N_b \cdot f_{clk}}{I\cdot(N^{cyc}_0+N^{cyc}_1)}
\label{eq:throughput}
\end{equation}
where $N_b$ is the number of decoded bits, $f_{clk}$ is the clock frequency, $I$ is the number of iterations, 
$N^{cyc}_0$ and $N^{cyc}_1$ are the number of clock cycles required to complete the interleaved 
and deinterleaved half iterations respectively. It is worth pointing out that $N_b=N$ for binary codes and $N_b=2N$ 
for double-binary codes. Results shown in the following sections have been obtained for $f_{clk}=200$ MHz and 
$I=8$ with the Turbo-NoC simulator \cite{turbo_NOC_download} and Synopsys Design Compiler 
for a 130 nm standard cell technology.

\subsection{ABR in NoC-based turbo decoder architectures}
\label{subsec:abr}

According to \cite{baghdadi_EL06} the throughput of an NOC-based turbo decoder can be increased by reducing the 
amount of data injected into the network. This approach is similar to well known early stopping criteria 
that are routinely used to both increase the throughput and reduce the power consumption in turbo decoder architectures 
\cite{chandrakasan_CICC10}. 
However, most of related works focus on frame-level early stopping criteria. On the contrary, bit-level/symbol-level 
early stopping criteria \cite{kim_CL06} take into account that the reliability of each bit/symbol in a frame converges at 
different speed. As a consequence, when the extrinsic information of a certain bit/symbol meets a proper reliability 
criterion, it is not necessary to further refine it. From an NoC-based turbo decoder perspective, this means 
that reliable $\lambda^{ext}_{i,j}[u]$ are no longer sent over the network. 
\begin{figure}
    \centering
      \psfrag{FL}[Bl][Bl][.55][0]{FL no ABR}
      \psfrag{FL THR=4}[Bl][Bl][.55][0]{FL $K=4$}
      \psfrag{FL THR=6}[Bl][Bl][.55][0]{FL $K=6$}
      \psfrag{FL THR=8}[Bl][Bl][.55][0]{FL $K=8$}
      \psfrag{FL THR=10}[Bl][Bl][.55][0]{FL $K=10$}
      \psfrag{RR}[Bl][Bl][.55][0]{RR no ABR}
      \psfrag{RR THR=4}[Bl][Bl][.55][0]{RR $K=4$}
      \psfrag{RR THR=6}[Bl][Bl][.55][0]{RR $K=6$}
      \psfrag{RR THR=8}[Bl][Bl][.55][0]{RR $K=8$}
      \psfrag{RR THR=10}[Bl][Bl][.55][0]{RR $K=10$}
      \includegraphics[width=\columnwidth]{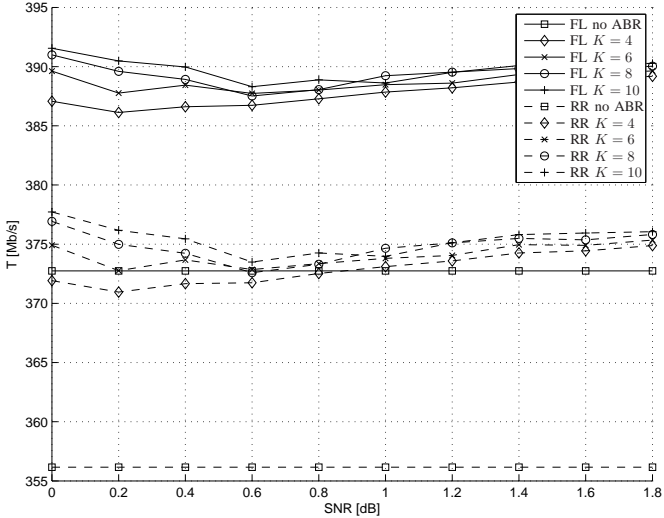}
\caption{Average throughput improvement at different SNR values of the HSDPA $N=5114$ turbo decoder with $K=4, 6, 8, 10$ 
on generalized Kautz networks $D=4$ and $P=64$}
\label{fig:ABR_hsdpa4}
\end{figure}

\subsection{HSDPA and 3GPP-LTE case of study}
\label{subsec:hsdpa_lte}

For binary turbo codes, as the ones employed in HSDPA and 3GPP-LTE standards, a simple ABR technique is obtained 
by fixing a threshold $K$ that is compared with $\delta = |\lambda^{ext}_{i,j}[u] - \lambda^{apr}_{i,j}[u]|$, namely  
if $\delta<K$, then $\lambda^{ext}_{i,j}[u]$ is not sent. The choice of $K$ depends not only on the specific code 
considered but also on the quantization parameters used to represent $\lambda^{ext}_{i,j}[u]$ and on the performance 
loss in terms of Bit-Error-Rate (BER) that can be accepted. In the following we consider $N=5114$ for HSDPA and 
$N=6144$ for 3GPP-LTE respectively. In both cases the extrinsic information is represented on eight bits whereas 
the intrinsic information is represented on six bits with three fractional bits. Both decoders perform eight iterations 
($I=8$) with $P=64$ using the Log-MAP algorithm \cite{robertson_ETT97} with a LUT-stored correction term. 
In Fig. \ref{fig:hsdpa} and \ref{fig:lte} we show the BER performance for the HSDPA and 3GPP-LTE codes 
respectively obtained by applying the ABR technique described in the previous paragraph with several 
values\footnote{Since we use three fractional bits for data representation the integer values of $K$ we 
considered correspond to 0.25, 0.75, 1 and so on.} for $K$. 
In particular, in Fig. \ref{fig:hsdpa} we show for the HSDPA code that when $K>10$ the performance 
worsens significantly. As an example, with $K=10$ there is a performance loss of less than 0.1 dB in the waterfall 
region and nearly ideal performance when the code floors. 
On the other hand, with $K=16$ the performance loss is of about 0.2 dB in the waterfall region and the code floor is 
shifted to higher SNR values of about 0.2 dB as well.
Similar results were observed for the 3GPP-LTE code, so, for the sake of clarity, in Fig. 
\ref{fig:lte} only results obtained with $K=4, 6, 8, 10$ are shown. 
For both cases we obtained the corresponding best and average bandwidth reduction at different SNR values through 
Monte Carlo simulations\footnote{The worst case corresponds to simulations where ABR is not applied}.
Experimental results show that the throughput increase is significant when there is a high load on the network ($R=1$) 
either using FL or RR routing algorithm. 
In particular, in Fig. \ref{fig:ABR_hsdpa2}, \ref{fig:ABR_hsdpa3} and \ref{fig:ABR_hsdpa4} we show the average 
throughput increase for the HSDPA turbo decoder for $D=2,3,4$ respectively with different values of $K$. 
As it can be observed 
when $R=1$ there is an average throughput increase, with respect to a decoder where ABR is not applied,  
that ranges from about 5 to 20 Mb/s for the HSDPA turbo decoder. 
Furthermore, we observed that in the best case there is a throughput increase of at least 60 Mb/s. 
On the other hand, when $R<1$ the average throughput improvement is at most of 5 Mb/s.
Similar results have been obtained for the LTE turbo decoder.

To complete the comparison, we show in Table \ref{tab:hsdpa_lte} the throughput/area results for the HSDPA and LTE 
cases respectively, where the results for the HSDPA case with ASP-FT routing algorithm and AP node architecture 
are taken from \cite{martina_MPMS11}. As it can be observed the significant throughput increase obtained with the
ABR technique on the FA node architecture when $R=1$ is paid as an area overhead with respect to the AP node architecture. 
However, as pointed out in \cite{martina_MPMS11}, the AP node architecture requires a large external memory to 
store the routing information. Moreover, the difference in terms of area between FA and AP node architectures reduces 
when $R<1$. In particular, as shown in Table \ref{tab:hsdpa_lte}, when $R=0.33$ with $P=8$ and $P=16$ the FA node 
architecture with the SSP-FL routing algorithm requires less area than the AP one.
\begin{table*}[t!]
  \centering
  \caption{Throughput [Mb/s]~-~area [mm$^2$] achieved with the HSDPA $N=5114$ and LTE $N=6144$ interleavers, 
with generalized Kautz topologies for $P \in \{8, 16, 32, 64\}$, $R \in \{0.33, 0.5, 1\}$, SSP-RR, SSP-FL and ASP-FT routing algorithms, no ABR} 
  \label{tab:hsdpa_lte}
   { \scriptsize
  \begin{tabular}{|c|c|c|c|c|c|c|c|c|c|}
\hline
& 	 & \multicolumn{4}{c|}{$D$=2, HSDPA} & \multicolumn{4}{c|}{$D$=2, LTE} \\
\hline
& 	 & $P$=8 & $P$=16 & $P$=32 & $P$=64  & $P$=8 & $P$=16 & $P$=32 & $P$=64 \\
\hline
\multirow{3}{*}{$R$=1.00} & SSP-RR (FA) & 54~-~3.13 & 74~-~5.29 & 105~-~7.36 & 159~-~9.71 
& 53~-~3.69 & 71~-~6.25 & 101~-~8.76 & 140~-~11.14 \\
& SSP-FL (FA) & 58~-~3.16 & 81~-~5.04 & 117~-~6.65 & 171~-~8.48 
& 54~-~3.88 & 75~-~5.93 & 109~-~7.79 & 151~-~9.74 \\
& ASP-FT (AP) & 58~-~1.53 & 81~-~2.51 & 117~-~3.40 & 171~-~4.51 
& 54~-~2.01 & 75~-~3.01 & 109~-~4.00 & 151~-~5.14 \\
\hline
\multirow{3}{*}{$R$=0.50} & SSP-RR (FA) & 46~-~0.59 & 72~-~1.71 & 101~-~4.07 & 142~-~6.64 
& 44~-~0.64 & 66~-~1.95 & 90~-~4.63 & 120~-~7.44 \\
& SSP-FL (FA) & 46~-~0.56 & 78~-~1.26 & 112~-~3.39 & 156~-~5.76 
& 44~-~0.61 & 72~-~1.37 & 100~-~3.79 & 131~-~6.31 \\
& ASP-FT (AP) \cite{martina_MPMS11} & 46~-~0.62 & 78~-~1.01 & 112~-~2.06 & 156~-~3.40 
& 44~-~0.71 & 72~-~1.13 & 100~-~2.34 & 131~-~3.72 \\
\hline
\multirow{3}{*}{$R$=0.33} & SSP-RR (FA) & 31~-~0.54 & 58~-~0.86 & 92~-~2.01 & 129~-~4.57 
& 29~-~0.59 & 52~-~0.90 & 79~-~2.25 & 102~-~4.84 \\
& SSP-FL (FA) & 31~-~0.53 & 58~-~0.81 & 101~-~1.56 & 140~-~3.68 
& 29~-~0.58 & 52~-~0.85 & 86~-~1.53 & 112~-~3.79 \\
& ASP-FT (AP) & 31~-~0.62 & 58~-~0.88 & 101~-~1.36 & 140~-~2.54 
& 29~-~0.72 & 52~-~0.98 & 86~-~1.44 & 112~-~2.69 \\
\hline
\hline
& 	 & \multicolumn{4}{c|}{$D$=3, HSDPA} & \multicolumn{4}{c|}{$D$=3, LTE} \\
\hline
& 	 & $P$=8 & $P$=16 & $P$=32 & $P$=64  & $P$=8 & $P$=16 & $P$=32 & $P$=64 \\
\hline
\multirow{3}{*}{$R$=1.00} & SSP-RR (FA) & 84~-~1.74 & 111~-~2.81 & 188~-~4.51 & 279~-~7.06 
& 75~-~1.89 & 107~-~3.29 & 161~-~5.07 & 232~-~8.08 \\
& SSP-FL (FA) & 90~-~1.00 & 126~-~2.54 & 194~-~4.44 & 291~-~6.82 
& 86~-~0.90 & 116~-~2.95 & 171~-~5.05 & 240~-~7.82 \\
& ASP-FT (AP) & 90~-~0.75 & 142~-~1.34 & 207~-~2.37 & 298~-~3.71 
& 86~-~0.76 & 132~-~1.50 & 185~-~2.69 & 254~-~4.18 \\
\hline
\multirow{3}{*}{$R$=0.50} & SSP-RR (FA) & 46~-~0.62 & 87~-~0.99 & 151~-~1.92 & 230~-~3.83 
& 44~-~0.66 & 78~-~1.00 & 128~-~1.81 & 178~-~3.96 \\
& SSP-FL (FA) & 46~-~0.61 & 86~-~0.96 & 152~-~1.77 & 237~-~3.41 
& 44~-~0.65 & 78~-~0.95 & 129~-~1.64 & 183~-~3.40 \\
& ASP-FT (AP) & 46~-~0.70 & 87~-~0.96 & 152~-~1.45 & 238~-~2.42 
& 44~-~0.79 & 78~-~1.04 & 129~-~1.48 & 186~-~2.45 \\
\hline
\multirow{3}{*}{$R$=0.33} & SSP-RR (FA) & 31~-~0.59 & 58~-~0.91 & 103~-~1.65 & 167~-~3.15 
& 29~-~0.64 & 52~-~0.94 & 87~-~1.59 & 129~-~3.06 \\
& SSP-FL (FA) & 31~-~0.59 & 58~-~0.90 & 103~-~1.62 & 167~-~3.02 
& 29~-~0.61 & 52~-~0.92 & 87~-~1.53 & 128~-~2.85 \\
& ASP-FT (AP) & 31~-~0.66 & 58~-~0.93 & 103~-~1.43 & 167~-~2.33 
& 29~-~0.74 & 52~-~1.00 & 87~-~1.46 & 129~-~2.35 \\
\hline
\hline
& 	 & \multicolumn{4}{c|}{$D$=4, HSDPA} & \multicolumn{4}{c|}{$D$=4, LTE} \\
\hline
& 	 & $P$=8 & $P$=16 & $P$=32 & $P$=64  & $P$=8 & $P$=16 & $P$=32 & $P$=64 \\
\hline
\multirow{3}{*}{$R$=1.00} & SSP-RR (FA) & 75~-~1.72 & 156~-~2.17 & 191~-~4.01 & 356~-~5.84 
& 66~-~1.98 & 133~-~2.40 & 167~-~4.40 & 301~-~6.03 \\
& SSP-FL (FA) & 83~-~1.31 & 163~-~1.63 & 199~-~3.89 & 372~-~5.51 
& 76~-~1.45 & 151~-~1.44 & 183~-~4.16 & 312~-~5.68 \\
& ASP-FT (AP) & 90~-~0.74 & 163~-~1.12 & 246~-~1.99 & 372~-~3.31 
& 86~-~0.78 & 151~-~1.12 & 217~-~2.10 & 312~-~3.45 \\
\hline
\multirow{3}{*}{$R$=0.50} & SSP-RR (FA) & 46~-~0.64 & 87~-~1.08 & 152~-~2.03 & 246~-~3.87 
& 44~-~0.67 & 79~-~1.04 & 130~-~1.87 & 192~-~3.39 \\
& SSP-FL (FA) & 46~-~0.62 & 87~-~1.05 & 152~-~1.94 & 245~-~3.80 
& 44~-~0.66 & 79~-~1.00 & 129~-~1.77 & 192~-~3.20 \\
& ASP-FT (AP) & 46~-~0.73 & 87~-~1.04 & 152~-~1.63 & 245~-~2.77 
& 44~-~0.84 & 79~-~1.13 & 130~-~1.65 & 192~-~2.66 \\
\hline
\multirow{3}{*}{$R$=0.33} & SSP-RR (FA) & 31~-~0.61 & 58~-~1.02 & 103~-~1.86 & 170~-~3.53 
& 29~-~0.65 & 53~-~1.01 & 87~-~1.75 & 130~-~3.16 \\
& SSP-FL (FA) & 31~-~0.61 & 58~-~0.99 & 104~-~1.82 & 170~-~3.51 
& 29~-~0.62 & 53~-~0.97 & 87~-~1.69 & 130~-~3.04 \\
& ASP-FT (AP) & 31~-~0.69 & 58~-~1.01 & 104~-~1.59 & 170~-~2.69 
& 29~-~0.77 & 53~-~1.05 & 87~-~1.60 & 130~-~2.61 \\
\hline
\end{tabular}
}
\end{table*}
\begin{figure}[t!]
  \centering
  \psfrag{symbol-level}[Bl][Bl][.6][0]{SL}
  \psfrag{bit-level}[Bl][Bl][.6][0]{BL}
  \psfrag{bit-level nx=6}[Bl][Bl][.6][0]{BL $n_{\xi}=6$}
  \psfrag{bit-level nx=5}[Bl][Bl][.6][0]{BL $n_{\xi}=5$}
  \psfrag{bit-level nx=4}[Bl][Bl][.6][0]{BL $n_{\xi}=4$}
  \psfrag{bit-level nx=4 THR=4}[Bl][Bl][.6][0]{BL $n_{\xi}=4$ $K=4$}
  \psfrag{bit-level nx=4 THR=6}[Bl][Bl][.6][0]{BL $n_{\xi}=4$ $K=6$}
  \includegraphics[width=\columnwidth]{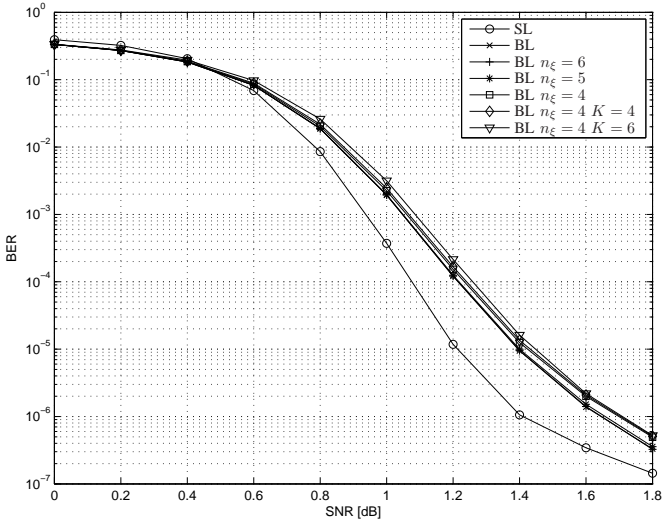}
\caption{BER performance of the WiMAX $N=1920$ turbo decoder with SL, BL, PFP representation and ABR technique, $K=4, 6$} 
\label{fig:wimax}
\end{figure}
\begin{figure}[t!]
    \centering
      \psfrag{FL}[Bl][Bl][.5][0]{FL no ABR}
      \psfrag{FL THR=4}[Bl][Bl][.5][0]{FL $K=4$}
      \psfrag{FL THR=6}[Bl][Bl][.5][0]{FL $K=6$}
      \psfrag{RR}[Bl][Bl][.5][0]{RR no ABR}
      \psfrag{RR THR=4}[Bl][Bl][.5][0]{RR $K=4$}
      \psfrag{RR THR=6}[Bl][Bl][.5][0]{RR $K=6$}
      \includegraphics[width=\columnwidth]{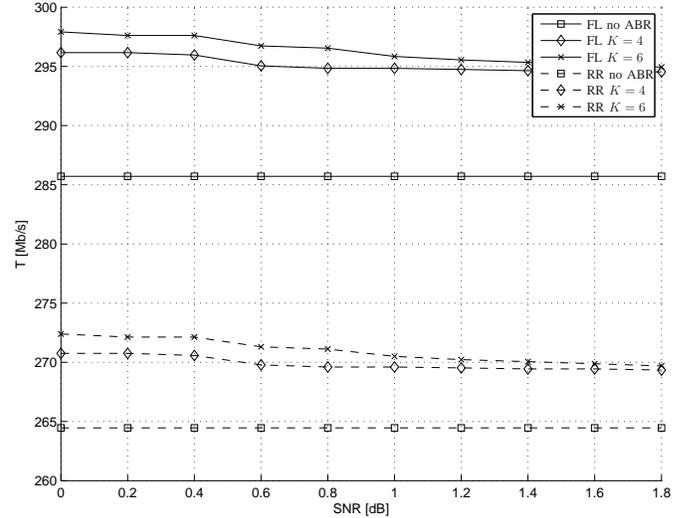}
\caption{Average throughput improvement at different SNR values of the WiMAX $N=1920$ turbo decoder with $K=4, 6$ 
on generalized Kautz networks $D=2$ and $P=64$}
\label{fig:ABR_wimax_1920_2}
\end{figure}

\subsection{WiMAX case of study} 

Simulation results shown in this section have been obtained with $N=1920$, as in \cite{kim_TCASII09}. 
Each component of the extrinsic information is represented on eight bits whereas the intrinsic information is 
represented on six bits with two fractional bits. The decoder performs eight iterations ($I=8$) with $P=64$ using the Max-Log-MAP 
algorithm \cite{robertson_ETT97}.
\begin{figure}[t!]
    \centering
      \psfrag{FL}[Bl][Bl][.5][0]{FL no ABR}
      \psfrag{FL THR=4}[Bl][Bl][.5][0]{FL $K=4$}
      \psfrag{FL THR=6}[Bl][Bl][.5][0]{FL $K=6$}
      \psfrag{RR}[Bl][Bl][.5][0]{RR no ABR}
      \psfrag{RR THR=4}[Bl][Bl][.5][0]{RR $K=4$}
      \psfrag{RR THR=6}[Bl][Bl][.5][0]{RR $K=6$}
      \includegraphics[width=\columnwidth]{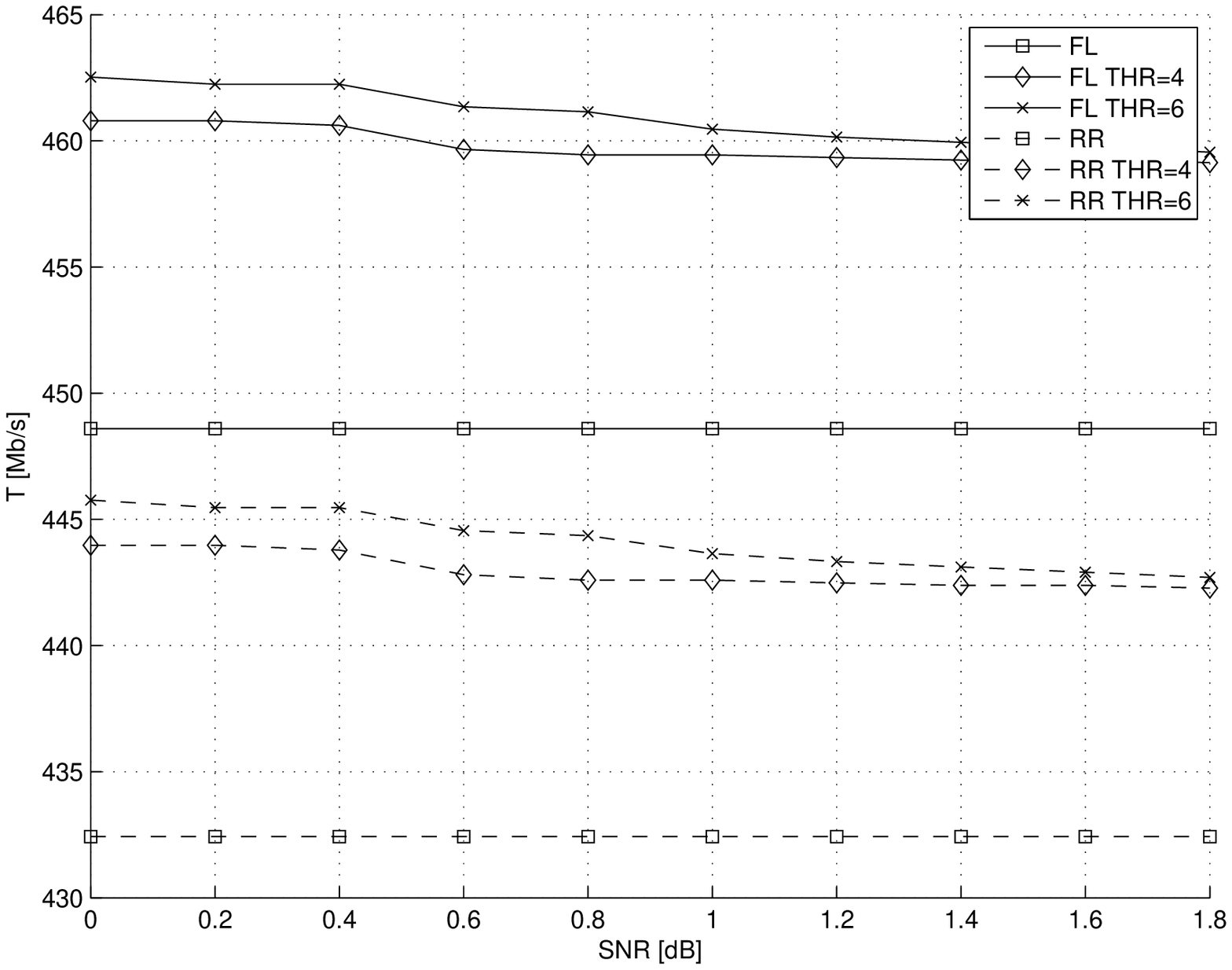}
\caption{Average throughput improvement at different SNR values of the WiMAX $N=1920$ turbo decoder with $K=4, 6$ 
on generalized Kautz networks $D=3$ and $P=64$}
\label{fig:ABR_wimax_1920_3}
\end{figure}

Since in binary turbo codes $\mathcal{U}=\{0,1\}$, the LLR of the extrinsic information is a scalar value. 
On the other hand, for double-binary turbo codes $\mathcal{U}=\{00, 01, 10, 11\}$, as a consequence  
$\lambda^{ext}_{i,j}[u]$ is an array containing three elements. In \cite{kim_TCASII09}, a bit level double-binary 
turbo decoder architecture is proposed to reduce the amount of memory to store the extrinsic information. 
The same idea is exploited in this work to reduce the area overhead of the NoC. Basically, a double-binary 
uncoded symbol $u$ can be represented as a couple of binary random variables $AB$. Then, with a slight abuse 
of notation, said $X$ a binary random variable, we denote $X=0$ with $\overline{X}$ and $X=1$ with $X$. 
Resorting to the Max-Log-MAP approximation we can convert Symbol-Level (SL) LLRs to Bit-Level (BL) LLRs as
\begin{eqnarray}
  \lambda^{ext}_{i,j}[A] & \simeq & \mu_{A} - \mu_{\overline{A}} \\
  \lambda^{ext}_{i,j}[B] & \simeq & \mu_{B} - \mu_{\overline{B}} 
\end{eqnarray}
where 
\begin{eqnarray}
\mu_{A} & = & \max\{\lambda^{ext}_{i,j}[A\overline{B}], \lambda^{ext}_{i,j}[AB]\} \\
\mu_{\overline{A}} & = & \max\{0, \lambda^{ext}_{i,j}[\overline{A}B] \} \\
\mu_{B} & = & \max\{\lambda^{ext}_{i,j}[\overline{A}B], \lambda^{ext}_{i,j}[AB]\} \\
\mu_{\overline{B}} & = & \max\{0, \lambda^{ext}_{i,j}[A\overline{B}] \}.
\end{eqnarray}
Similarly, we can convert BL LLRs to SL LLRs with the following approximations.
\begin{enumerate}
\item $\lambda^{ext}_{i,j}[A] \ge 0$ and $\lambda^{ext}_{i,j}[B] \ge 0$
\begin{eqnarray}
\lambda^{ext}_{i,j}[A\overline{B}] & \simeq & \mu_{AB} - \lambda^{ext}_{i,j}[B] \\
\lambda^{ext}_{i,j}[\overline{A}B] & \simeq & \mu_{AB} - \lambda^{ext}_{i,j}[A] \\
\lambda^{ext}_{i,j}[AB]            & \simeq & \mu_{AB}
\end{eqnarray}
\item $\lambda^{ext}_{i,j}[A] \ge 0$ and $\lambda^{ext}_{i,j}[B] < 0$
\begin{eqnarray}
  \lambda^{ext}_{i,j}[A\overline{B}] & \simeq & \lambda^{ext}_{i,j}[A] \\
  \lambda^{ext}_{i,j}[\overline{A}B] & \simeq & 0 \\
  \lambda^{ext}_{i,j}[AB]            & \simeq & \lambda^{ext}_{i,j}[A]+\lambda^{ext}_{i,j}[B] 
\end{eqnarray}
\item $\lambda^{ext}_{i,j}[A] < 0$ and $\lambda^{ext}_{i,j}[B] \ge 0$
\begin{eqnarray}
  \lambda^{ext}_{i,j}[A\overline{B}] & \simeq & 0 \\
  \lambda^{ext}_{i,j}[\overline{A}B] & \simeq & \lambda^{ext}_{i,j}[B] \\
  \lambda^{ext}_{i,j}[AB]            & \simeq & \lambda^{ext}_{i,j}[A]+\lambda^{ext}_{i,j}[B]
\end{eqnarray}
\item $\lambda^{ext}_{i,j}[A] < 0$ and $\lambda^{ext}_{i,j}[B] < 0$
\begin{eqnarray}
  \lambda^{ext}_{i,j}[A\overline{B}] & \simeq & \lambda^{ext}_{i,j}[A] \\
  \lambda^{ext}_{i,j}[\overline{A}B] & \simeq & \lambda^{ext}_{i,j}[B] \\
  \lambda^{ext}_{i,j}[AB]            & \simeq & \lambda^{ext}_{i,j}[A]+\lambda^{ext}_{i,j}[B] - \mu_{AB}
\end{eqnarray}
\end{enumerate}
where 
\begin{equation}
\mu_{AB} = \max \{\lambda^{ext}_{i,j}[A], \lambda^{ext}_{i,j}[B]\}
\end{equation}
For further details on bit to symbol and symbol to bit conversion the reader can refer to \cite{kim_TCASII09}.
\begin{table*}[t!]
  \centering
  \caption{Throughput [Mb/s]~-~area SL [mm$^2$]~-~area BL [mm$^2$], PFP achieved with the WiMAX $N=1920$ interleaver, 
with generalized Kautz topologies for $P \in \{8, 16, 32, 64\}$, $R \in \{0.33, 0.5, 1\}$, SSP-RR, SSP-FL and ASP-FT routing algorithms, no ABR} 
  \label{tab:wimax}
   { \scriptsize
  \begin{tabular}{|c|c|c|c|c|c|}
\hline
& 	 & \multicolumn{4}{c|}{$D$=2} \\
\hline
& 	 & $P$=8 & $P$=16 & $P$=32 & $P$=64 \\
\hline
\multirow{3}{*}{$R$=1.00} & SSP-RR (FA) & 104~-~2.15~-~1.46 & 138~-~3.61~-~2.43 & 195~-~5.16~-~3.51 & 264~-~6.97~-~4.85 \\
& SSP-FL (FA) & 105~-~2.17~-~1.47 & 144~-~3.40~-~2.30 & 208~-~4.57~-~3.13 & 285~-~6.11~-~4.29 \\
& ASP-FT (AP) & 105~-~1.62~-~0.92 & 144~-~2.54~-~1.43 & 208~-~3.42~-~1.99 & 285~-~4.61~-~2.79 \\
\hline
\multirow{3}{*}{$R$=0.50} & SSP-RR (FA) & 86~-~0.47~-~0.38 & 127~-~1.48~-~1.07 & 176~-~3.20~-~2.26 & 231~-~5.33~-~3.80 \\
& SSP-FL (FA) & 86~-~0.42~-~0.35 & 134~-~1.19~-~0.88 & 187~-~2.74~-~1.96 & 246~-~4.60~-~3.33 \\
& ASP-FT (AP) & 86~-~0.42~-~0.35 & 134~-~1.00~-~0.69 & 187~-~2.15~-~1.37 & 246~-~3.56~-~2.29 \\
\hline
\multirow{3}{*}{$R$=0.33} & SSP-RR (FA) & 58~-~0.39~-~0.33 & 102~-~0.78~-~0.62 & 153~-~1.94~-~1.45 & 199~-~4.05~-~2.98 \\
& SSP-FL (FA) & 58~-~0.36~-~0.31 & 103~-~0.69~-~0.57 & 161~-~1.55~-~1.20 & 209~-~3.41~-~2.56 \\
& ASP-FT (AP) & 58~-~0.38~-~0.33 & 103~-~0.67~-~0.54 & 161~-~1.33~-~0.99 & 209~-~2.73~-~1.89 \\
\hline
\hline
& 	 & \multicolumn{4}{c|}{$D$=3} \\
\hline
& 	 & $P$=8 & $P$=16 & $P$=32 & $P$=64 \\
\hline
\multirow{3}{*}{$R$=1.00} & SSP-RR (FA) & 148~-~1.07~-~0.77 & 204~-~2.19~-~1.53 & 306~-~3.59~-~2.53 & 432~-~5.70~-~4.08 \\
& SSP-FL (FA) & 165~-~0.69~-~0.53 & 218~-~2.10~-~1.48 & 328~-~3.39~-~2.41 & 448~-~5.39~-~3.89 \\
& ASP-FT (AP) & 165~-~0.59~-~0.43 & 249~-~1.34~-~0.86 & 344~-~2.57~-~1.60 & 452~-~4.07~-~2.59 \\
\hline
\multirow{3}{*}{$R$=0.50} & SSP-RR (FA) & 87~-~0.47~-~0.38 & 152~-~0.90~-~0.71 & 243~-~1.80~-~1.38 & 333~-~3.87~-~2.91 \\
& SSP-FL (FA) & 87~-~0.45~-~0.37 & 152~-~0.85~-~0.68 & 242~-~1.68~-~1.31 & 334~-~3.45~-~2.64 \\
& ASP-FT (AP) & 87~-~0.47~-~0.39 & 152~-~0.80~-~0.63 & 244~-~1.43~-~1.07 & 338~-~2.75~-~1.96 \\
\hline
\multirow{3}{*}{$R$=0.33} & SSP-RR (FA) & 58~-~0.44~-~0.37 & 103~-~0.84~-~0.67 & 168~-~1.64~-~1.28 & 243~-~3.27~-~2.53 \\
& SSP-FL (FA) & 58~-~0.42~-~0.35 & 103~-~0.80~-~0.64 & 167~-~1.57~-~1.23 & 243~-~3.10~-~2.41 \\
& ASP-FT (AP) & 58~-~0.44~-~0.37 & 103~-~0.76~-~0.60 & 168~-~1.38~-~1.05 & 243~-~2.57~-~1.89 \\
\hline
\hline
& 	 & \multicolumn{4}{c|}{$D$=4} \\
\hline
& 	 & $P$=8 & $P$=16 & $P$=32 & $P$=64 \\
\hline
\multirow{3}{*}{$R$=1.00} & SSP-RR (FA) & 135~-~1.24~-~0.88 & 253~-~1.79~-~1.29 & 323~-~3.41~-~2.45 & 513~-~5.38~-~3.93 \\
& SSP-FL (FA) & 153~-~0.94~-~0.69 & 279~-~1.41~-~1.05 & 344~-~3.21~-~2.33 & 533~-~5.09~-~3.76 \\
& ASP-FT (AP) & 166~-~0.63~-~0.46 & 279~-~1.15~-~0.80 & 393~-~2.28~-~1.51 & 533~-~3.99~-~2.66 \\
\hline
\multirow{3}{*}{$R$=0.50} & SSP-RR (FA) & 87~-~0.50~-~0.41 & 155~-~0.92~-~0.73 & 247~-~1.98~-~1.53 & 354~-~3.83~-~2.94 \\
& SSP-FL (FA) & 87~-~0.48~-~0.39 & 155~-~0.90~-~0.72 & 248~-~1.89~-~1.47 & 356~-~3.70~-~2.84 \\
& ASP-FT (AP) & 87~-~0.52~-~0.43 & 155~-~0.87~-~0.69 & 249~-~1.66~-~1.24 & 356~-~3.12~-~2.27 \\
\hline
\multirow{3}{*}{$R$=0.33} & SSP-RR (FA) & 58~-~0.47~-~0.39 & 104~-~0.92~-~0.73 & 169~-~1.85~-~1.45 & 248~-~3.61~-~2.80 \\
& SSP-FL (FA) & 58~-~0.44~-~0.36 & 104~-~0.88~-~0.70 & 169~-~1.75~-~1.37 & 248~-~3.45~-~2.67 \\
& ASP-FT (AP) & 58~-~0.48~-~0.40 & 104~-~0.84~-~0.66 & 169~-~1.57~-~1.19 & 248~-~2.97~-~2.19 \\
\hline
\end{tabular}
}
\end{table*}

The use of BL LLRs introduces a BER performance loss of about 0.2 dB (see Fig. \ref{fig:wimax}), 
but it reduces the data width of one third with respect to SL LLRs,
as the payload of each packet contains $\lambda^{ext}_{i,j}[A]$ and $\lambda^{ext}_{i,j}[B]$ 
instead of $\lambda^{ext}_{i,j}[u]$. 
To further reduce the data width we applied to BL LLRs the Pseudo-Floating-Point (PFP) representation suggested 
in \cite{lee_VTC08}.
As highlighted also in \cite{vogt_EL00,masera_ISTC08} the most significant bits of the extrinsic information 
play an important role in the decoding procedure. 
Indeed, the basic idea is to analyze the binary representation of $\lambda^{ext}_{i,j}[A]$ and $\lambda^{ext}_{i,j}[B]$ 
(as 2's complement values) from the most significant bit to the least significant bit and to detect the first 
zero-one or one-zero transition, which represents the starting bit of the extrinsic information significand. 
We denote the significand as $\xi$ and the number of bits that prefix $\xi$ are coded as a shift index $\sigma$. 
Thus, for each couple $\lambda^{ext}_{i,j}[A]$, $\lambda^{ext}_{i,j}[B]$ we obtain $\xi_{i,j}[A]$, $\xi_{i,j}[B]$, 
$\sigma_{i,j}[A]$ and $\sigma_{i,j}[B]$. 
Then, according with \cite{lee_VTC08}, we impose $\sigma_{i,j}=\min\{\sigma_{i,j}[A], \sigma_{i,j}[B]\}$. 
Said $n_{\lambda}$, $n_{\xi}$ and $n_{\sigma}$ the number of bits to represent $\lambda$, $\xi$ and $\sigma$ 
respectively we obtain 
\begin{eqnarray}
\tilde{\xi}_{i,j}[A] & = & \lambda^{ext}_{i,j}[A] >> (n_{\lambda} - n_{\xi} - \sigma_{i,j}) \\
\tilde{\xi}_{i,j}[B] & = & \lambda^{ext}_{i,j}[B] >> (n_{\lambda} - n_{\xi} - \sigma_{i,j}) 
\end{eqnarray}
where $>>$ stands for arithmetic right shift. As a consequence, the payload of each packet sent on the network 
now contains $\tilde{\xi}_{i,j}[A]$, $\tilde{\xi}_{i,j}[B]$ and $\sigma_{i,j}$ instead of $\lambda^{ext}_{i,j}[u]$.
\begin{figure}[h!]
    \centering
      \psfrag{FL}[Bl][Bl][.5][0]{FL no ABR}
      \psfrag{FL THR=4}[Bl][Bl][.5][0]{FL $K=4$}
      \psfrag{FL THR=6}[Bl][Bl][.5][0]{FL $K=6$}
      \psfrag{RR}[Bl][Bl][.5][0]{RR no ABR}
      \psfrag{RR THR=4}[Bl][Bl][.5][0]{RR $K=4$}
      \psfrag{RR THR=6}[Bl][Bl][.5][0]{RR $K=6$}
      \includegraphics[width=\columnwidth]{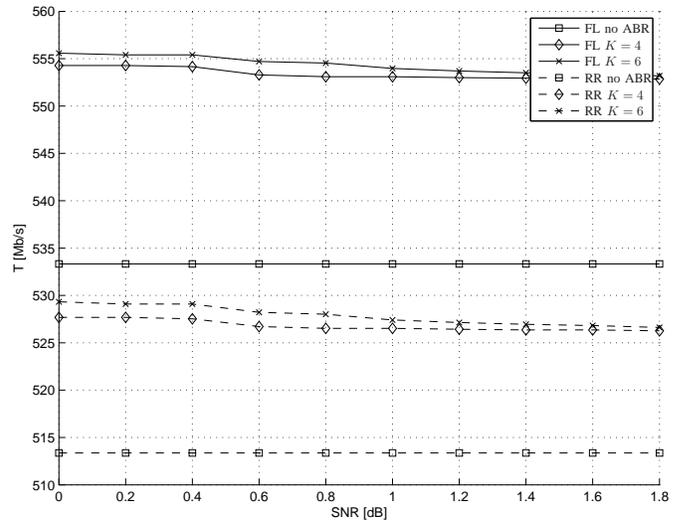}
\caption{Average throughput improvement at different SNR values of the WiMAX $N=1920$ turbo decoder with $K=4, 6$ 
on generalized Kautz networks $D=4$ and $P=64$}
\label{fig:ABR_wimax_1920_4}
\end{figure}

As stated in the first paragraph of this section $n_{\lambda}=8$. Thus, said $n_d$ the number of bits 
devoted to represent the extrinsic information in the payload we have: i) $n_d=3n_{\lambda}=24$ for $\lambda^{ext}_{i,j}[u]$ and
ii) $n_d=2n_{\lambda}=16$ for $\lambda^{ext}_{i,j}[A]$ and $\lambda^{ext}_{i,j}[B]$.
If we impose $n_{\xi}=4$, we obtain $\sigma_{i,j} \le 4$ and so $n_{\sigma}=3$, leading to $n_d=2n_{\xi}+n_{\sigma}=11$ that is less 
than half the value of $n_d$ for $\lambda^{ext}_{i,j}[u]$. As shown in Fig. \ref{fig:wimax} the BER performance loss 
of BL, PFP LLR representation, is nearly the same as the fixed point BL one.
In Table \ref{tab:wimax} the throughput and area results obtained by using SL and BL, PFP LLR 
representation are shown for generalized Kautz topologies. 
As it can be observed, the area decrease as a function of $n_d$ is not linear, however, 
it becomes particularly interesting when $R=1$. 
As an example, with $R=1$, $D=4$ and $P=64$ there is an area saving of up to the 40\%.

The techniques described in the previous paragraphs are all aimed at reducing the area of the NoC-based turbo 
decoder. Furthermore, the ABR technique described in section \ref{subsec:abr} can be used to improve the throughput 
as well. In order to limit the BER performance loss introduced by the ABR technique, we employ the SL
reliability criterion proposed in \cite{baghdadi_EL06} but we send BL, PFP extrinsic information
when the criterion is not met. 
The ABR technique we used is summarized in Algorithm \ref{algo:baghdadi} and can be summarized as follows:
\begin{algorithm}
  \caption{SL reliability criterion proposed in \cite{baghdadi_EL06} }
  \label{algo:baghdadi}
  \begin{algorithmic}[1]
    \STATE $\vartheta^{apr}_{i,j} \leftarrow \max \{\lambda^{apr}_{i,j}[u]\}$ 
    \STATE $\varrho^{apr}_{i,j} \leftarrow \max \{\lambda^{apr}_{i,j}[u] \setminus \vartheta^{apr}_{i,j}\}$ 
    \STATE $\vartheta^{ext}_{i,j} \leftarrow \max \{\lambda^{ext}_{i,j}[u]\}$
    \STATE $\varrho^{ext}_{i,j} \leftarrow \max \{\lambda^{ext}_{i,j}[u] \setminus \vartheta^{ext}_{i,j}\}$ 
    \STATE $\Delta^{apr}_{i,j} \leftarrow |\vartheta^{apr}_{i,j}-\varrho^{apr}_{i,j}|$ 
    \STATE $\Delta^{ext}_{i,j} \leftarrow |\vartheta^{ext}_{i,j}-\varrho^{ext}_{i,j}|$ 
    \STATE $\Phi_{i,j} \leftarrow |\Delta^{ext}_{i,j}-\Delta^{apr}_{i,j}|$
    \IF{$\Phi_{i,j} < K$}
    \STATE do not send any packet
    \ELSE
    \STATE send $\tilde{\xi}_{i,j}[A]$, $\tilde{\xi}_{i,j}[B]$, $\sigma_{i,j}$
    \ENDIF
  \end{algorithmic}
\end{algorithm}
said $\vartheta^{apr}_{i,j}$, $\varrho^{apr}_{i,j}$ and 
$\vartheta^{ext}_{i,j}$, $\varrho^{ext}_{i,j}$ 
the first and the second maximum values in $\lambda^{apr}_{i,j}[u]$ and $\lambda^{ext}_{i,j}[u]$ respectively, we 
compute $\Delta^{apr}_{i,j}=|\vartheta^{apr}_{i,j}-\varrho^{apr}_{i,j}|$ and 
$ \Delta^{ext}_{i,j}=|\vartheta^{ext}_{i,j}-\varrho^{ext}_{i,j}|$; finally, we compare 
$\Phi_{i,j}=|\Delta^{ext}_{i,j}-\Delta^{apr}_{i,j}|$ with the threshold $K$.

As shown in Fig. \ref{fig:wimax} the BER performance loss introduced by the ABR technique is negligible. 
Moreover, as shown in Fig. \ref{fig:ABR_wimax_1920_2}, \ref{fig:ABR_wimax_1920_3} and \ref{fig:ABR_wimax_1920_4} 
when $R=1$ the ABR technique induces an average throughput increase of about 5 to 20 Mb/s. 
Similarly to the binary codes in the best case the throughput improvement is at least of more than 40 Mb/s, whereas 
when $R<1$ the average throughput improvement is at most of 5 Mb/s.

\section{Conclusions}
\label{sec:concl}
In this work ABR techniques have been exploited to improve the throughput of NoC-based turbo decoder architectures. 
When the load of the network is high the average throughput is improved of about 5 to 20 Mb/s and in the best case 
the throughput is increased of more than 60 Mb/s and 40 Mb/s for binary and double-binary codes respectively. 
Moreover, the area required to support double-binary codes has been 
significantly reduced (up to more than the 40\%) by applying BL, PFP representation of the extrinsic information with 
a BER performance loss of about 0.2 dB.

\bibliographystyle{IEEEtran}
\bibliography{biblio}

\begin{thebibliography}{10}
\providecommand{\url}[1]{#1}
\csname url@samestyle\endcsname
\providecommand{\newblock}{\relax}
\providecommand{\bibinfo}[2]{#2}
\providecommand{\BIBentrySTDinterwordspacing}{\spaceskip=0pt\relax}
\providecommand{\BIBentryALTinterwordstretchfactor}{4}
\providecommand{\BIBentryALTinterwordspacing}{\spaceskip=\fontdimen2\font plus
\BIBentryALTinterwordstretchfactor\fontdimen3\font minus
  \fontdimen4\font\relax}
\providecommand{\BIBforeignlanguage}[2]{{%
\expandafter\ifx\csname l@#1\endcsname\relax
\typeout{** WARNING: IEEEtran.bst: No hyphenation pattern has been}%
\typeout{** loaded for the language `#1'. Using the pattern for}%
\typeout{** the default language instead.}%
\else
\language=\csname l@#1\endcsname
\fi
#2}}
\providecommand{\BIBdecl}{\relax}
\BIBdecl

\bibitem{802-16}
``{IEEE} {S}td 802.16, part 16: air interface for fixed broadband wireless
  access systems,'' Oct. 2004.

\bibitem{lte}
``{TS 36.212 v8.0.0: Multiplexing and Channel Coding (FDD) (Release 8)},''
  2007-09.

\bibitem{berrou_ICC93}
C.~Berrou, A.~Glavieux, and P.~Thitimajshima, ``Near {S}hannon limit error
  correcting coding and decoding: {T}urbo codes,'' in \emph{IEEE International
  Conference on Communications}, 1993, pp. 1064--1070.

\bibitem{gallager_TrIT62}
R.~G. Gallager, ``Low density parity check codes,'' \emph{IRE Transactions on
  Information Theory}, vol. IT-8, no.~1, pp. 21--28, Jan 1962.

\bibitem{montorsi_IEEEProc07}
E.~Boutillon, C.~Douillard, and G.~Montorsi, ``Iterative decoding of
  concatenated convolutional codes: Implementation issues,'' \emph{Proceedings
  of the IEEE}, vol.~95, no.~6, pp. 1201--1227, Jun 2007.

\bibitem{boutillon_TCOM07}
F.~Guilloud, E.~Boutillon, J.~Tousch, and J.-L. Danger, ``Generic description
  and synthesis of {LDPC} decoders,'' \emph{IEEE Transactions on
  Communications}, vol.~55, no.~11, pp. 2084--2091, Nov 2007.

\bibitem{martina_MPMS11}
M.~Martina, G.~Masera, H.~Moussa, and A.~Baghdadi, ``On chip interconnects for
  multiprocessor turbo decoding architectures,'' \emph{Elsevier Microprocessors
  and Microsystems}, vol.~35, no.~2, pp. 167--181, Mar 2011.

\bibitem{polydoros_PIMRC08}
A.~Polydoros, ``Algorithmic aspects of radio flexibility,'' in \emph{IEEE
  International Symposium on Personal, Indoor and Mobile Communications}, 2008,
  pp. 1--5.

\bibitem{wehn_TVLSI08}
T.~Vogt and N.~Wehn, ``Reconfigurable {ASIP} for convolutional and turbo
  decoding in an {SDR} environment,'' \emph{IEEE Transactions on VLSI},
  vol.~16, no.~10, pp. 1309--1320, Oct 2008.

\bibitem{bougard_ICT08}
B.~Bougard, R.~Priewasser, L.~V. der Perre, and M.~Huemer,
  ``Algorithm-architecture co-design of a multi-standard {FEC} decoder
  {ASIP},'' in \emph{ICT Mobile Summit Conference}, 2008.

\bibitem{martina_TCASII08}
M.~Martina, M.~Nicola, and G.~Masera, ``A flexible {UMTS-WiMax} turbo decoder
  architecture,'' \emph{IEEE Transactions on Circuits and Systems II}, vol.~55,
  no.~4, pp. 369--373, Apr 2008.

\bibitem{kim_CICC09}
J.~H. Kim and I.~C. Park, ``A unified parallel radix-4 turbo decoder for mobile
  {WiMAX and 3GPP-LTE},'' in \emph{IEEE Custom Integrated Circuits Conference},
  2009, pp. 487--490.

\bibitem{baghdadi_TVLSI09}
O.~Muller, A.~Baghdadi, and M.~Jezequel, ``From parallelism levels to a
  multi-{ASIP} architecture for turbo decoding,'' \emph{IEEE Transactions on
  VLSI}, vol.~17, no.~1, pp. 92--102, Jan 2009.

\bibitem{baghdadi_soc10}
P.~Reddy, R.~Alkhayat, F.~Clermidy, A.~Baghdadi, and M.~Jezequel, ``Power
  consumption analysis and energy efficient optimization for turbo decoder
  implementation,'' in \emph{International Symposium on Sytem-on-Chip}, 2010,
  pp. 12--17.

\bibitem{wehn_ICCS10}
T.~Ilnseher, M.~May, and N.~Wehn, ``A multi-mode {3GPP-LTE/HSDPA} turbo
  decoder,'' in \emph{IEEE International Conference on Communication Systems},
  201, pp. 336--340.

\bibitem{vacca_DSD09}
F.~Vacca, H.~Moussa, A.~Baghdadi, and G.~Masera, ``Flexible architectures for
  {LDPC} decoders based on network on chip paradigm,'' in \emph{Euromicro
  Conference on Digital System Design}, 2009, pp. 582--589.

\bibitem{wehn_iscas05}
C.~Neeb, M.~J. Thul, and N.~Wehn, ``Network-on-chip-centric approach to
  interleaving in high throughput channel decoders,'' in \emph{IEEE
  International Symposium on Circuits and Systems}, 2005, pp. 1766--1769.

\bibitem{moussa_date07}
H.~Moussa, O.~Muller, A.~Baghdadi, and M.~Jezequel, ``Butterfly and
  {B}enes-based on-chip communication networks for multiprocessor turbo
  decoding,'' in \emph{Design, Automation and Test in Europe Conference and
  Exhibition}, 2007, pp. 654--659.

\bibitem{moussa_iscas08}
H.~Moussa, A.~Baghdadi, and M.~Jezequel, ``Binary de {B}ruijn interconnection
  network for a flexible {LDPC}/turbo decoder,'' in \emph{IEEE International
  Symposium on Circuits and Systems}, 2008, pp. 97--100.

\bibitem{martina_TCASI10}
M.~Martina and G.~Masera, ``Turbo {NOC}: a framework for the design of network
  on chip based turbo decoder architectures,'' \emph{IEEE Transactions on
  Circuits and Systems I}, vol.~57, no.~10, pp. 2776--2789, Oct 2010.

\bibitem{baghdadi_EL06}
O.~Muller, A.~Baghdadi, and M.~Jezequel, ``Bandwidth reduction of extrinsic
  information exchange in turbo decoding,'' \emph{IET Electronics Letters},
  vol.~42, no.~19, pp. 1104--1105, Sep 2006.

\bibitem{lee_VTC08}
S.~M. Park, J.~Kwak, and K.~Lee, ``Extrinsic information memory reduced
  architecture for non-binary turbo decoder implementation,'' in \emph{IEEE
  Vehicular Technology Conference}, 2008, pp. 539--543.

\bibitem{kim_TCASII09}
J.~H. Kim and I.~C. Park, ``Bit-level extrinsic information exchange method for
  double-binary turbo codes,'' \emph{IEEE Transactions on Circuits and Systems
  II}, vol.~56, no.~1, pp. 81--85, Jan 2009.

\bibitem{berrou_ITW01}
C.~Berrou, M.~Jezequel, C.~Douillard, and S.~Kerouedan, ``The advantages of
  non-binary turbo codes,'' in \emph{IEEE Information Theory Workshop}, 2001,
  pp. 61--63.

\bibitem{bahl_TrIT94}
L.~Bahl, J.~Cocke, F.~Jelinek, and J.~Raviv, ``Optimal decoding of linear codes
  for minimizing symbol error rate,'' \emph{IEEE Transactions on Information
  Theory}, vol.~20, no.~3, pp. 284--287, Mar 1974.

\bibitem{robertson_ICC95}
P.~Robertson, E.~Villebrun, and P.~Hoeher, ``A comparison of optimal and
  sub-optimal {MAP} decoding algorithms operating in the {L}og domain,'' in
  \emph{IEEE ICC}, 1995, pp. 1009--1013.

\bibitem{robertson_ETT97}
P.~Robertson, P.~Hoeher, and E.~Villebrun, ``Optimal and sub-optimal maximum a
  posteriori algorithms suitable for turbo decoding,'' \emph{European
  Transactions on Telecommunications}, vol.~8, no.~2, pp. 119--125, Mar-Apr
  1997.

\bibitem{martina_CL09}
S.~Papaharalabos, P.~T. Mathiopoulos, G.~Masera, and M.~Martina, ``On optimal
  and near-optimal turbo decoding using generalized $\max^*$ operator,''
  \emph{IEEE Communications Letters}, vol.~13, no.~7, pp. 522--524, Jul 2009.

\bibitem{hsdpa}
``http://www.3gpp2.org.''

\bibitem{wang_TCAS207}
Z.~Wang and Q.~Li, ``Very low-complexity hardware interleaver for turbo
  decoding,'' \emph{IEEE Transactions on Circuits and Systems II}, vol.~54,
  no.~7, pp. 636--640, Jul 2007.

\bibitem{martina_CL08}
M.~Martina, M.~Nicola, and G.~Masera, ``Hardware design of a low complexity,
  parallel interleaver for wimax duo-binary turbo decoding,'' \emph{IEEE
  Communications Letters}, vol.~12, no.~11, pp. 846--848, Nov 2008.

\bibitem{imase_TC81}
M.~Imase and M.~Itoh, ``Design to minimize diameter on building-block
  network,'' \emph{IEEE Transactions on Computers}, vol.~30, no.~6, pp.
  439--442, Jun 1981.

\bibitem{imase_TC83}
------, ``A design for directed graphs with minimum diameter,'' \emph{IEEE
  Transactions on Computers}, vol.~32, no.~8, pp. 782--784, Aug 1983.

\bibitem{turbo_NOC_download}
M.~Martina, ``{T}urbo {NOC}: {N}etwork {O}n {C}hip based turbo decoder
  architectures,'' downloadable at www.vlsilab.polito.it/$\sim$martina.

\bibitem{chandrakasan_CICC10}
C.~C. Cheng, Y.~M. Tsai, L.~G. Chen, and A.~P. Chandrakasan, ``A 0.077 to 0.168
  nj/bit/iteration scalable {3GPP LTE} turbo decoder with an adaptive sub-block
  parallel scheme and an embedded {DVFS} engine,'' in \emph{IEEE Custom
  Integrated Circuits Conference}, 2010, pp. 1--4.

\bibitem{kim_CL06}
D.~H. Kim and S.~W. Kim, ``Bit-level stopping of turbo decoding,'' \emph{IEEE
  Communications Letters}, vol.~10, no.~3, pp. 183--185, Mar 2006.

\bibitem{vogt_EL00}
J.~Vogt, J.~Ertel, and A.~Finger, ``Reducing bit width of extrinsic memory in
  turbo decoder realisations,'' \emph{IEE Electronics Letters}, vol.~36,
  no.~20, pp. 1714--1716, Sep 2000.

\bibitem{masera_ISTC08}
A.~Singh, E.~Boutillon, and G.~Masera, ``Bit-width optimization of extrinsic
  information in turbo decoder,'' in \emph{International Symposium on Turbo
  Codes \& Related Topics}, 2008, pp. 134--138.

\end{thebibliography}

\end{document}